\documentclass{article}

\usepackage{arxiv}

\usepackage[utf8]{inputenc} 
\usepackage[T1]{fontenc}    
\usepackage{hyperref}       
\usepackage{url}            
\usepackage{booktabs}       
\usepackage{amsfonts}       
\usepackage{nicefrac}       
\usepackage{microtype}   
\usepackage{graphicx}
\usepackage{doi}
\usepackage{multicol}

\title{Scaling laws and a general theory for the growth of public companies}

\author{Jiang Zhang \\
	School of Systems Science \\
	Beijing Normal University\\
	Beijing, China \\
	\texttt{zhangjiang@bnu.edu.cn} \\
	\And
	Christopher P. Kempes \\
	The Santa Fe Institute\\
	Santa Fe, NM, USA \\
	\texttt{ckempes@santafe.edu} \\
	\And
	Marcus J. Hamilton \\
	Department of Anthropology \\
	The University of Texas at San Antonio\\
	San Antonio, TX, USA \\
	\texttt{marcus.hamilton@utsa.edu} \\
	\And
	Ruyi Tao \\
	School of Systems Science \\
	Beijing Normal University\\
	Beijing, China \\
	\texttt{zhangjiang@bnu.edu.cn} \\
	\And
	Geoffrey B. West \\
	The Santa Fe Institute\\
	Santa Fe, NM, USA \\
	\texttt{gbw@santafe.edu} \\
}

\renewcommand{\shorttitle}{Company Growth}

\begin{document}
\maketitle

\begin{abstract}
Publicly traded companies are fundamental units of contemporary economies and markets and are important mechanisms through which humans interact with their environments. Understanding the general properties that underlie the processes of their growth  has long been of interest, yet fundamental debates about the effects of firm size on growth have persisted. Here we develop a scaling framework that focuses on company size as the critical feature determining a variety of tradeoffs, and use this to reveal novel systematic behavior across the diversity of publicly-traded companies. Using a large database of 31,553 US companies over nearly 70 years, and 3,160 Chinese companies over 24 year, we show how the dynamics of companies expressed as scaling relationships leads to a quantitative, analytic theory for their growth. This theory produces several predictions that are in good agreement with data for both the US and China, whose markets have strikingly different histories and underlying structures.  In both cases sales scale sublinearly with assets and exhibit nearly identical exponents leading, surprisingly and nontrivially, to assets that grow as a power law in time rather than exponentially, as often assumed. On the other hand, liabilities scale linearly in the US (exponent of $1.0$) but superlinearly in China (exponent of $1.09$). We show that such small differences in scaling exponents can have a significant impact on the character and long-term evolution of growth trajectories. These results illustrate that while companies are part of a larger class of growth phenomena driven by incomes and costs that scale with size, they are unique in that they grow following a temporal power-function which sets them apart from  organisms, cities, nations, and markets, whose growth over time is often exponential. The perspective developed here highlights novel dynamics in the scaling and growth of human economies.
\end{abstract}

\section{Introduction}\label{sec1}

Companies are major drivers of economic growth, employment, and technological innovation, and markets are the set of interactions through which resources, services, and wealth are generated and redistributed at all scales of society. Consequently, understanding the mechanisms that drive company growth in both the public and private sectors has been a primary goal of much research in economics and finance \cite{coase1937nature, penrose1959theory, williamson1993nature, carroll2004demography}. 
Furthermore, a quantitative understanding of the regularities that govern the dynamics of companies is central to understanding the dynamics of the modern world and addressing its long-term sustainability. In this paper, we investigate the interplay of growth, size, and age in companies to derive a general growth equation inspired by data on publicly-traded companies in the US and China, which are then used to test its many predictions \cite{davis1992gross, davis1998job, haltiwanger2013creates, cooper1994initial, moreno2021family, birch1987job, dixon2012firm, malecki2018entrepreneurs}.

From an ecological perspective, companies are individual entities that interact, compete, and cooperate with each other for finite resources within complex environments and with complex interconnections. From this perspective individual companies are complex adaptive systems governed by the same types of regularities, rules, and dynamics governing individual organisms and their collectives ranging from bacteria to hunter-gatherer groups and cities. Many of the most successful laws discovered for complex adaptive systems concern the regular patterns exhibited by large ensembles of entities. Indeed, companies have already been shown to display some of the statistical signatures of similar types of laws such as an exponential distribution of lifespans~\cite{daepp2015mortality}, and Zipfian rank-abundance distribution of sizes with  Laplacian fluctuations (e.g. \cite{axtell2001zipf,stanley1995zipf,stanley1996scaling}).

Much like organisms and cities, the continuing challenge of adaptability, evolvability and growth in response to competitive forces requires companies to be {\it scalable}. Despite their extraordinary complexity and diversity, many of the most fundamental characteristics of organisms, such as metabolic rates, growth rates, lifespans and genome lengths, scale with size in a surprisingly simple and general fashion ranging from cells to ecosystems \cite{west2005origin}. Similarly, many socio-economic and physical characteristics of cities, including wages, patents, crime, and road lengths, scale systematically with size~\cite{bettencourt2007growth}. These scaling laws are often the consequence of the generic optimization principles of competitive entities under constraints \cite{kempes2019scales}, a concept that is expected to generalise to companies. Given the success of the scaling perspective, we extend the growth paradigm developed within this framework \cite{west2001general,kempes2012growth} to over 30,000 publicly traded US companies and over 3,000 Chinese companies.  

The mechanisms traditionally suggested for understanding companies fall into three broad inter-related categories that are often treated separately. In the language developed for understanding organisms and cities, these are: (i) Organizational structure, which is the network system that conveys information, resources and capital to support, sustain and grow the company; (ii) Minimization of transaction costs to enhance economies of scale and maximize profits;  (iii) Competitive forces inherent in the ecology of the market place. Properties such as these likely underlie the origin of the scaling laws \cite{coase1937nature,penrose1959theory,williamson1993nature,carroll2004demography}. 

Historically, perhaps the most prominent theory of public company growth is Gibrat's hypothesis of the law of proportional effect \cite{gibrat1931inegalites}. This proposes that firm growth is dominated by random fluctuations so the distribution of company sizes converges to lognormal \cite{shimizu2018history}, and their proportional growth rate is independent of size. Empirically, however, this is not supported by data \cite{santarelli2006gibrat}. Among the many extensions of Gibrat's statistical hypothesis are a class of models that assume that companies are composed of sub-units that grow according to a generalized preferential attachment model \cite{fu2005growth,buldyrev2007generalized,riccaboni2008size} that can account for the tent-like fluctuations in  growth rates~\cite{stanley1996scaling,perline2006volatility}. It is also worth noting that significant differences may exist between private and publicly traded companies even for such broad statistical regularities (e.g. \cite{perline2006volatility}). 

In contrast to these statistical approaches to growth, our framework focuses on the {\it mechanisms} that drive it, namely, the flows and utilisation of capital derived from profits and financing flowing through the company, and which are governed by the scaling laws. We show how these lead to a generalized equation for company growth whose predictions are in good quantitative agreement with data. 

\section{Theoretical Framework and Results}\label{sec2}

\subsection*{Cash flows, stores, and the growth of companies}
\begin{figure*}[ht!]
    \centering
    \includegraphics[width=1.0\textwidth]{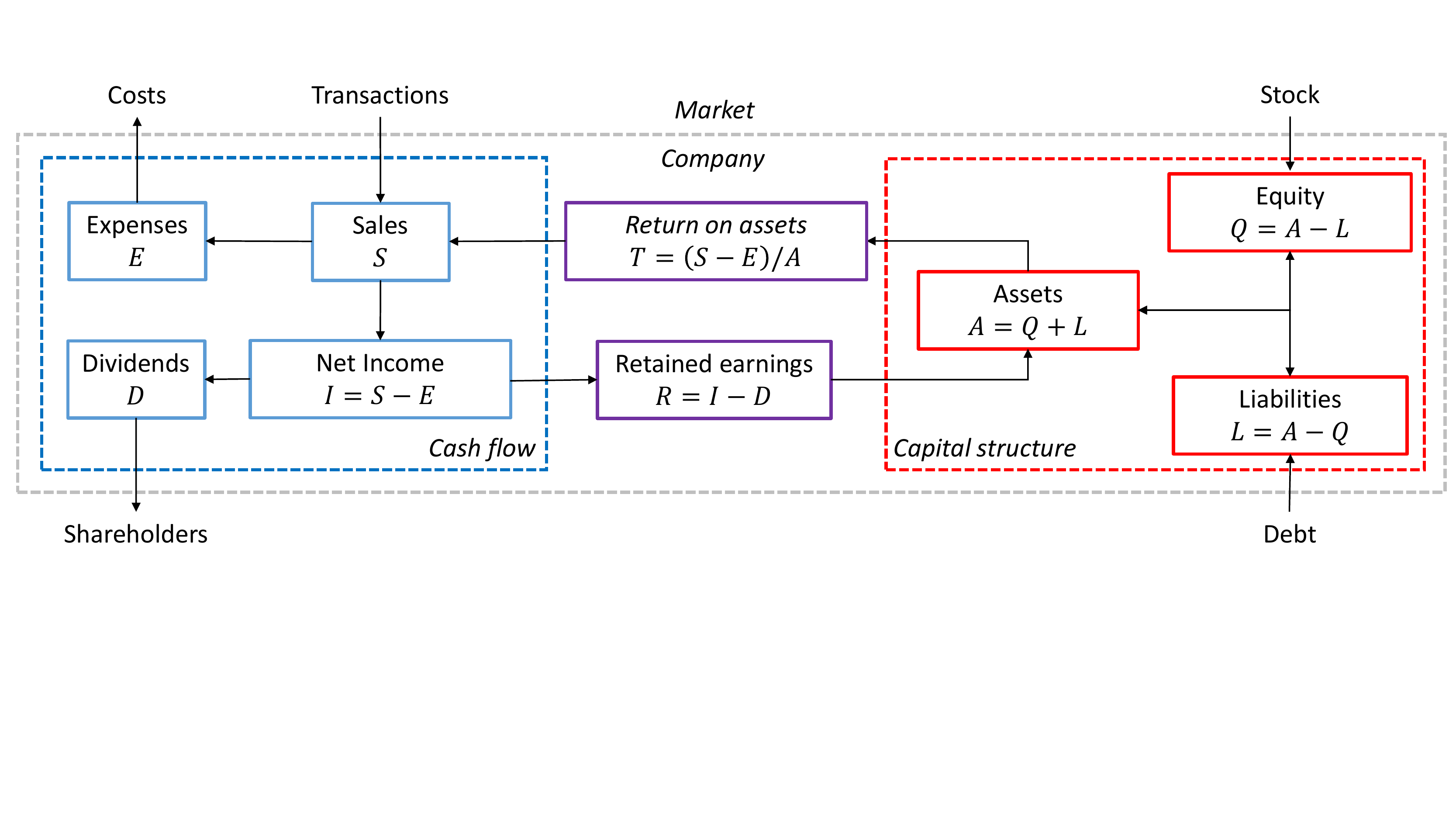}
    \caption{The cash flow and capital structure of financial variables as represented in the COMPUSTAT database. Cash flow is the flux of financial resources through a company generated by sales. Sales are generated from transactions of goods and services outside the company within the market. Net income (i.e., net profit/loss) is the difference between sales and expenses (i.e., costs released back into the market). Once dividends have been paid to shareholders the remainder constitutes the retained earnings which accumulate as assets. The capital structure of a company is the structure of assets which are apportioned into equity (i.e., issued stock) and liabilities (i.e., issued debt). The ability of a company to deploy assets to generate cash flow constitutes the return on assets and is a common measure of the financial efficiency of a company. All the quantities scale with assets. The exponents shown in the figure are for the US companies.}
    \label{fig:stockflows}
\end{figure*}

The size and growth of a company is ultimately determined by flows and stores through the system. This is illustrated in Fig. \ref{fig:stockflows} which shows the inter-relationship between the various quantities conventionally used to characterize the financial state of a company, as represented in the COMPUSTAT database. We classify these variables into three categories: 1) income related quantities, which include sales, net income, and gross profit; 2) cost related quantities, which include cost of sales, total taxes, operating costs, and R\&D; and 3) other size-related quantities, such as the number of employees, assets, liabilities, and the availability of cash. The incoming flows consist of two parts, sales and finance, whereas the outgoing flows represent expenses. We identify the amount of financial resources stored at a given time as a company's assets which we use as the measure for its size. 

\subsection*{Note on assets as a cumulative measure of company size}

While common measures of company size are, for example, the number of employees or sales,  here we use assets as the natural metric to capture size. Assets are resources with economic value that are accrued by a company in the expectation they can be used for future economic benefit. They are accumulated (or lost) over the lifetime of a company and are a direct measure of that company's historic ability to accumulate economic resources. Accounting standards are somewhat flexible and the reporting of assets on financial statements can be manipulated by companies in order to either enhance their profile to investors, or dissuade potential acquirers. However, overt manipulation of financial statements is illegal, and while it can be difficult to detect, it is unlikely that the flexibility (or illegality) of accounting standards ranges over scales of orders of magnitude that we are concerned with here.

\subsection*{Scaling the income, cost, and size of companies}

Companies are, on average, self-similar and scale-invariant if, at any time $t$, they obey power function scaling laws (e.g. \cite{shalit1977measurement}) for their various measurable properties, $X$,  such as their sales, net income, profit or total liabilities. This is expressed mathematically as:

\begin{equation}
\label{eq:scaling}
    X(A)=c_XA^{\beta_X}
\end{equation}
where $A$ are their assets, $c_X$ a normalization constant and $\beta_X$ a scaling exponent; both $c_X$ and $\beta_X$ are time and scale invariant, but can depend on the sector or market in which companies operate. 

\begin{figure*}
\center
\includegraphics[width=1.0\textwidth]{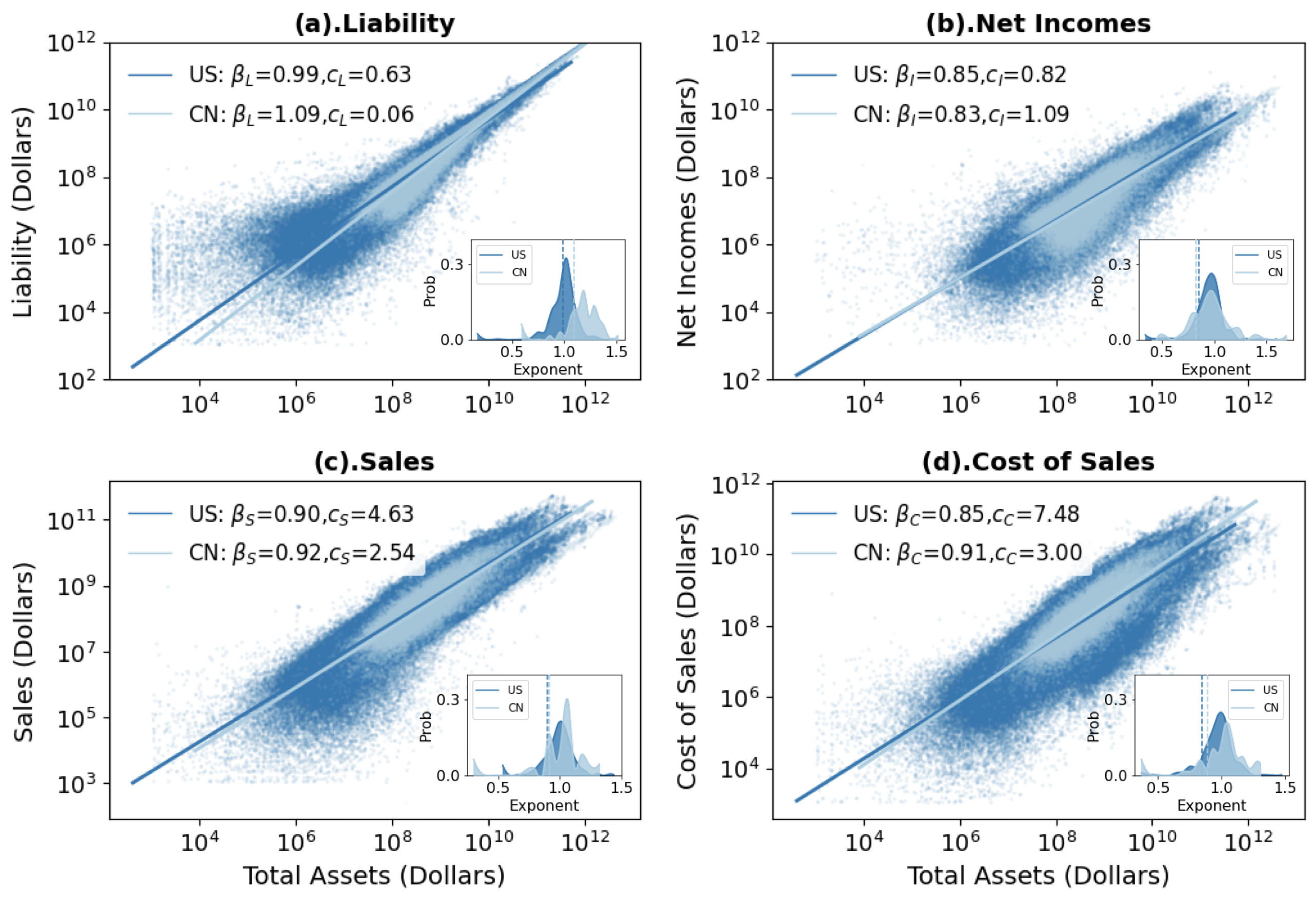}
\caption{The scaling laws of total liabilities (\textbf{a}), net income (\textbf{b}), annual sales  (\textbf{c}), and annual cost of sales (\textbf{d}) as a power function of total assets for all companies over all years and for all economic sectors. Dark blue represents American companies and light blue represents Chinese companies. The units of all the variables are US dollars. The solid lines are fitted scaling functions using ordinary least square (OLS) regression over all sectors. The inset shows the distributions of the fitted exponents (with a Linear Mixed-effect Model, LMM; see methods) for 443 different industrial sectors for the US and 62 sectors for China. Further statistical results are available in Section S2 of the Supplementary Materials.}
\label{fig:scalings}
\end{figure*}

We test these ideas and estimate the suite of parameters $c_X$ and $\beta_X$ using a dataset of 31,553 publicly-traded companies in US markets covering 1950-2018, and a dataset of 3,160 publicly-traded companies in Chinese markets covering 1996-2020. Taking the logarithm of Eq. (\ref{eq:scaling}) yields the linear form $\ln X=\ln c_X+\beta_X \ln A$, so a plot of $\ln X$ vs. $\ln A$ should yield a sequence of straight lines whose slopes are the exponents $\beta_X$. As can readily be seen in Fig. \ref{fig:scalings}, all income and cost variables do indeed exhibit significant power function scaling, albeit with fluctuations that we also discuss in detail.  

Table \ref{tab:scaling} summarises the scaling exponents, $\beta_X$, for selected variables averaged across the entire market as well as for the two countries. While there is some variation in their values, they are almost all sub-linear (that is, less than $1$) indicating that, regardless of the sector, assets grow systematically faster than any company metric including sales, cost of sales and net income, as shown in Fig. \ref{fig:scalings}. A notable exception is liabilities for which the exponent $\beta_L$ converges to $1$ for the US market, but is larger than $1$ for the Chinese market, indicating that they scale linearly (US) or super-linearly (China) with company size; this characteristic plays a central role in determining the nature of growth trajectories, which we address in much greater detail later. The inset panels show the distributions of exponents across different sectors within each market; the few outlying ones are likely due to sectors with only a small sampling of firms (see section S.1 in SI for details). These regularities and the close clustering of exponents for a given variable across different sectors suggests that similar underlying dynamics and network organizational principles are at play across companies regardless of their size, age or economic sector. An important observation that plays a crucial role in determining their growth trajectory is that income-related variables have comparatively larger exponents than cost-related ones. A broad variety of previous work has discussed the size dependence of financial features like sales and growth rates (e.g. \cite{shalit1977measurement}). Here our focus is to relate all of the financial features of a company to a common assessment of size to reveal their interconnection and, through a budget, to derive their dynamics and growth trajectories, and in turn to predict their sensitivity to changes in exponents. 

\subsection*{Growth Equation}

\begin{figure*}[ht!]
\centering
\includegraphics[width=1.0\textwidth]{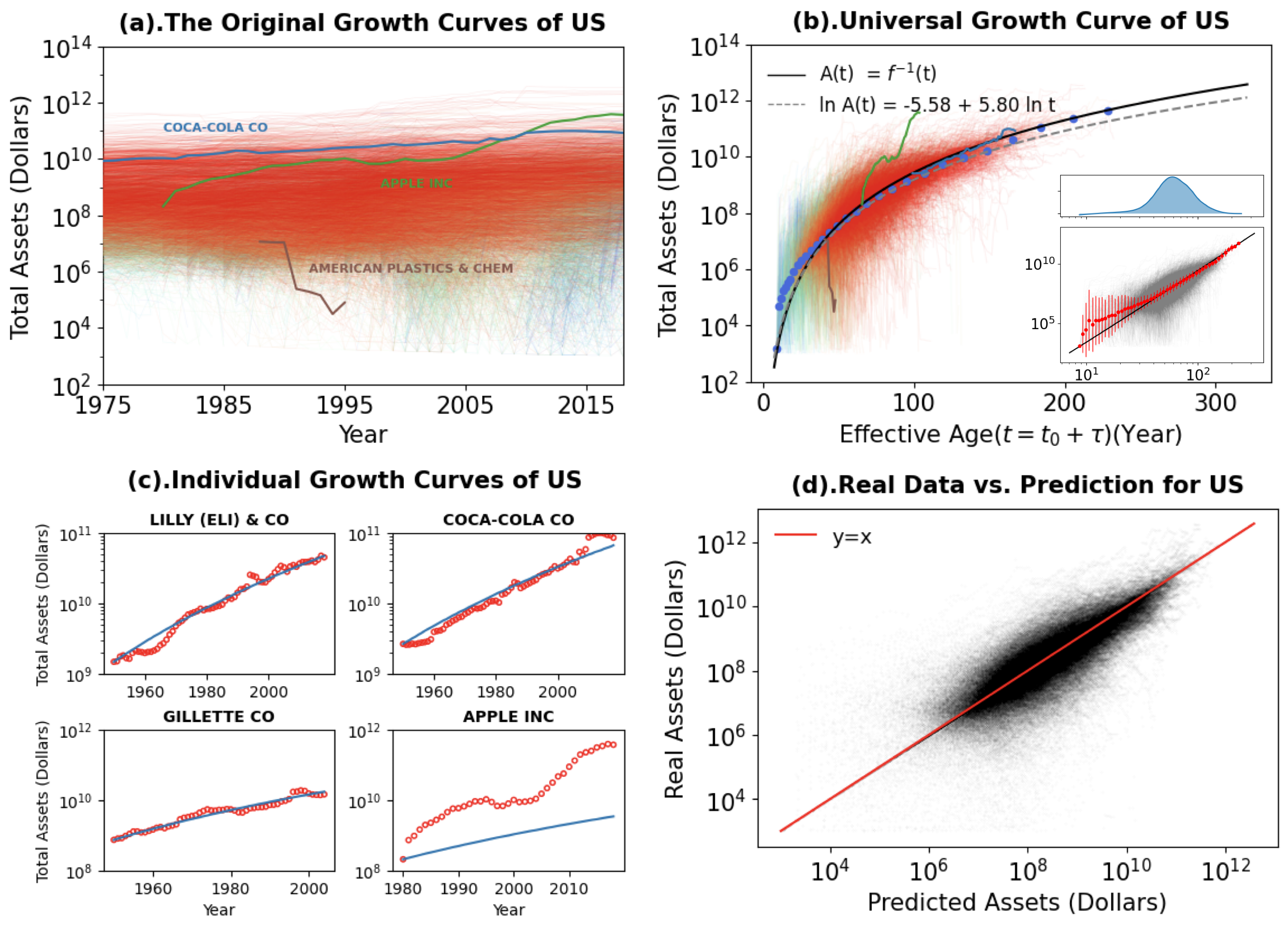}
\caption{\textbf{Publicly traded market of the United States}. The growth trajectories of all 31,553 companies in terms of total assets over time and the generalized growth predictions. (\textbf{a}) The raw growth trajectories of all companies in our dataset from 1975 to 2018. Each curve represents a single company. Three representative companies are highlighted: Apple Inc. (green), Coca-Cola Co. (blue), and American Plastics \& Chem. (brown). These companies were chosen as they reflect companies out-performing predictions (Apple Inc.), meeting predictions (Coca-Cola Co.), or under-performing expectations (American Plastics \& Chem.).  
(\textbf{b}) The predicted generalized growth curves and the translated growth trajectories of each company based on the estimated $t_0$, highlighting that companies grow as a power function of time. The inset shows the same data (gray points) along with the log-binned averages (the red points) and the same  generalized growth curve (black line). The error bar indicates the standard deviation of the data and the histogram provides the distribution all of the data. The predicted power function exponent $\gamma $ and the coefficient $\ln c$ are also shown. (\textbf{c}) Observed growth trajectories (red circles) and their corresponding predicted growth (blue solid lines) obtained by solving Eq.~(\ref{eq:powerlawgrowth}) with the observed initial size for four selected companies. The time is the natural year. (\textbf{d}) The comparison of assets from observed growth data along with the growth predicted by Eq.~(\ref{eq:powerlawgrowth}) showing excellent agreement, as manifested by the red straight line having a slope of 1. Across panels (a)-(b) companies are consistently colored according to their lifespans in the data (defined as the difference between their first and last appearance in the data set, which does not necessarily correspond with their actual births and deaths). Other forms of the growth equation are also shown to compare in S5 of the SI.}
\label{fig:growthcurves}
\end{figure*}

The detailed growth trajectory of any individual company is a complex response of its particular internal structure and the external conditions it confronts. Fig. \ref{fig:growthcurves}(a) shows the diversity of these trajectories for the 31,553 publicly-traded companies in th US market since 1975. Larger, more mature companies tend to grow slowly following the average growth of the market, whereas smaller, younger ones tend to grow more rapidly but with greater fluctuations. Nevertheless, underlying the stochasticity of these individual growth trajectories are, at any time $t$, size-dependent regularities captured by scaling laws. We show how these time invariant constraints, reflecting the common underlying dynamics shared by all companies, can be used to derive an explicit growth equation that determines their generic growth trajectories. Indeed, we show that the wide diversity of these trajectories can be reduced to a single size-dependent generalized growth process. 

If $S(t)$ is the annual sales of a company (with the unit dollars/year) and $F(t)$ is the investment raised annually in financial markets (with the unit dollars/year and see SI S3.1), then the total amount of money flowing into the company during a time interval $\Delta t$ (year) is $[S(t) + F(t)]\Delta t$ (dollars). As illustrated in Fig.~\ref{fig:stockflows}, this total amount of incoming money is partially used to maintain the company by paying its  expenses, $E(t)\Delta t$ (dollars), during this time interval, where $E(t)$ (dollars/year) is the total annual expenses which include wages, taxes, cost of sales and equipment, etc. The remainder fuels the company's growth by contributing to an increase in its assets, $\Delta A(t)$ (dollars);
consequently, $[S(t) + F(t)]\Delta t = E(t)\Delta t + \Delta A(t)$, so
\begin{equation}
\label{eq:balance 1}
   \Delta A(t) = [S(t) + F(t) - E(t)] \Delta t
\end{equation}
Finances raised from the market are the dominant contribution to the increase in a company's total liabilities, $L(t)$, thus,
$F(t)\Delta t \approx \Delta L(t)$(see SI S4.1 for the verification of this approximation). Furthermore, the difference between sales ($S$) and total expenses ($E$) is net income ($I$ in dollars/year), i.e., $S(t) - E(t) = I(t)$, so Eq.~(\ref{eq:balance 1}) can be re-expressed as
$ \Delta A(t) = I(t) \Delta t + \Delta L(t)$. Taking the limit $\Delta t \rightarrow 0$ gives

\begin{equation}
\label{eq:balance 2}
\frac{dA}{dt} = I + \frac{dL}{dt}
\end{equation}

This balance equation is exact at any time, $t$. However, quantities such as sales, expenses and income, are not reported instantaneously; instead, they are reported annually, effectively making time discrete rather than continuous with one year being the minimum time unit. Consequently, when using results based on Eq.~(\ref{eq:balance 2}) in which $\Delta t \rightarrow 0$, and comparing them with data which are reported only annually, we inevitably introduce some unknown, though relatively small, error(see SI S4.1 for the details). 

With scaling laws in mind and noting that $\frac{dL}{dt}=(\frac{dL}{dA})( \frac{dA}{dt}$), it is prudent to consider $I$ and $L$ as functions of company size, $A(t)$, in which case Eq.~(\ref{eq:balance 2}) can be re-expressed as 
\begin{equation}
\label{eq:generalgrowth}
    \frac{dA}{dt}=\frac{I(A)}{1-\frac{dL(A)}{dA}}.
\end{equation}
Consequently, there are two separate strategies for ensuring positive growth (i.e., $dA/dt > 0$): either (i) having positive income ($I > 0$) coupled with the differential debt ratio $dL/dA < 1$; or (ii) having negative income ($I < 0$) coupled with the differential debt ratio $dL/dA > 1$. Note that, in general, it is the {\it differential} debt ratio, $dL/dA$, rather than the conventional debt ratio, $k(A) = L/A$, that is a determining factor for positive growth. In what follows we restrict the discussion and corresponding data to cases where net income is positive since these dominate the market; furthermore, on average, the inclusion of short-term losses does not appreciably affect long-term development of companies, nor the value of the exponents. We relegate the discussion of negative income to section S2.4 of the SI.

Introducing the scaling relationships $I(A)= c_IA^{\beta_I}$ and $L(A)= c_LA^{\beta_L}$ into 
Eq.~(\ref{eq:generalgrowth}) leads to

\begin{equation}
    \label{eq:powerlawgrowth}
    \frac{dA}{dt}=\frac{c_IA^{\beta_I}}{1-c_L\beta_LA^{\beta_L-1}}
\end{equation}
This can be straightforwardly integrated to give
\begin{equation}
    \label{eq:integratingequation}
\frac{A^{1-\beta_I}}{c_I(1-\beta_I)}\left[1-\left(\frac{1-\beta_I}{\beta_L-\beta_I}\right)c_L\beta_LA^{\beta_L-1}\right]\equiv f(A) = t
\end{equation}
where we have imposed the initial condition $A(0)=0$ at $t=0$ (see S3.3 of SI for the complete expression, and S3.2 for the solution when $A(0) \neq 0$). In general, it is not possible to invert Eq.~(\ref {eq:integratingequation}) to obtain an analytic expression for $A(t)=f^{-1}(t)$. 

However, we saw in 
Fig.~\ref{fig:scalings} and Table~\ref{tab:scaling} that, $\beta_L \rightarrow 1$ for the US market (i.e., liabilities are a constant fraction of assets), in which case Eq.~(\ref {eq:integratingequation}) reduces to a simple tractable analytic solution for the averaged growth trajectory of companies: 
\begin{equation}
    \label{eq:simplesolution}
    A(t) = \left[\frac{c_I(1-\beta_I)}{1-c_L}t\right]^{1/(1-\beta_I)}
\end{equation}
More generally, the exact solution to Eq.~(\ref{eq:integratingequation}) for asymptotically large $t$ and $A(t)$, i.e, for large mature companies, is given by $A(t) = \left[{c_I(1-\beta_I)} t\right]^{1/(1-\beta_I)}$, provided $\beta_L < 1$, a condition satisfied for US companies. Therefore, in general, the theory predicts that companies, on average, grow following a simple power function of time:
\begin{equation}
    \label{eq:simplesolution2}
    A(t) \approx c t^{\gamma}
\end{equation}
with both $c$ and $\gamma$ predicted by the theory. We refer to Eq.~(\ref{eq:powerlawgrowth}) as the generalized growth equation and Eq.~(\ref{eq:simplesolution2}) as the generalized growth curve with $t$ being the effective age.
Since $\beta_I$ is close to $1$ ($\approx 0.85$), the leading order solution, Eq.~(\ref{eq:simplesolution}), predicts that $A(t)$ grows rapidly with a relatively large exponent, $\gamma = 1/(1-\beta_I) \approx 6.7$, though {\it not} as fast as an exponential. It is straightforward to calculate the leading corrections to Eq.~(\ref {eq:simplesolution}) by expanding $A(t)$ in Eq.~(\ref {eq:integratingequation}) in a Taylor series around $\beta_L=1$. The predominant behaviour is still a power function as in Eq.~(\ref{eq:simplesolution2}), but with a logarithmic correction leading to deviations for smaller companies; the explicit formulae and details are presented in SI S3.4. The refined prediction gives $\gamma \approx 5.8$.

The power function relationship (Eq.~(\ref{eq:simplesolution2})) is confirmed by the data as shown in a semi-logarithmic plot (Fig.~\ref{fig:growthcurves}(b)), in which the leading order prediction is a logarithmic curve (the solid line). Shown in the inset of Fig.~\ref{fig:growthcurves}(b) is a log-log plot of the data, demonstrating that the leading order prediction is a straight line with the slope $\gamma \approx 5.8$. Importantly, 
Fig.~\ref{fig:growthcurves} also confirms the deviation from the strict power predicted by the full theory for smaller companies, Eq.~(\ref{eq:integratingequation}).  
Compared to the large diversity of individual company growth trajectories across all sectors shown in Fig.~\ref{fig:growthcurves}(a) the collapse of the data to a single generalized growth curve is illustrated by Fig.~\ref{fig:growthcurves}(b) . 

In confronting the generalized growth model from Eq.~(\ref{eq:simplesolution2}) with data it is important to note that there are many companies whose first year of operation predates the reported time period. To deal with this we can infer an effective initial year, $t_0^i$, that best aligns the data with the generalized curve. This is done by solving for the initial time as $t_0^i=\left({A_0^i}/{c}\right)^{1/\gamma}$ using the initial assets, $A_0^i$, and then plotting the generalized growth curve as
\begin{equation}
    \label{eq:individual}
    \hat{A}^i(\tau)=c\cdot\left(\tau+t_0^i\right)^{\gamma}.
\end{equation}
Here, $\tau$ is the years we observe the company in the data starting from the inferred age of $t_0^i$, and  $\hat{A}^i(\tau)$ is the predicted total assets of a company in year $\tau$. Fits of $t_0^i$ for each individual company reveal an impressive generalized curve in Fig.~\ref{fig:growthcurves}(b). We also validate the method for inferring $t_0$ by a group of companies with randomly generated initial startup year in S4.2 of SI.

A more direct test of our theory is to solve Eq.~(\ref{eq:powerlawgrowth}) using the observed initial size of each individual company (see S3.2 of SI) in which case we find an impressively good fit to the data without employing any free parameters other than the cross-company scaling exponents and constants. Fig.~\ref{fig:growthcurves}(c) shows four examples for comparison between theoretical prediction and real growth trajectories of individual companies. Three show excellent agreement between prediction and data, while the fourth example is an outlier showing over-performance of a company (Apple) relative to the idealised prediction. Notice, however, that this trajectory still approximates a power function, albeit with a higher exponent. In general, Fig.~\ref{fig:growthcurves}(d) shows good agreements between the predicted value of the assets compared to their actual values for all companies in all years. This is equivalent to the generalized growth curve because, as discussed in S3.3 of the SI, ideally each individual curve can be regarded as a part of the generalized growth curve. 

\subsection*{Testing the generality of the theory in China}
A crucial test of the generality of our theory is provided by confronting its predictions with data from both the US and Chinese markets whose histories and underlying systems are strikingly different.
Our theoretical framework connects the form of growth trajectories with the scaling exponents of financial features of companies so changes in these exponents, such as between US and China, should lead to measurable differences predicted by the growth equation. To explore this idea we turn to 3,160 publicly-traded companies in the Chinese market since 1996. 

\begin{figure*}[ht!]
\includegraphics[width=1.0\textwidth]{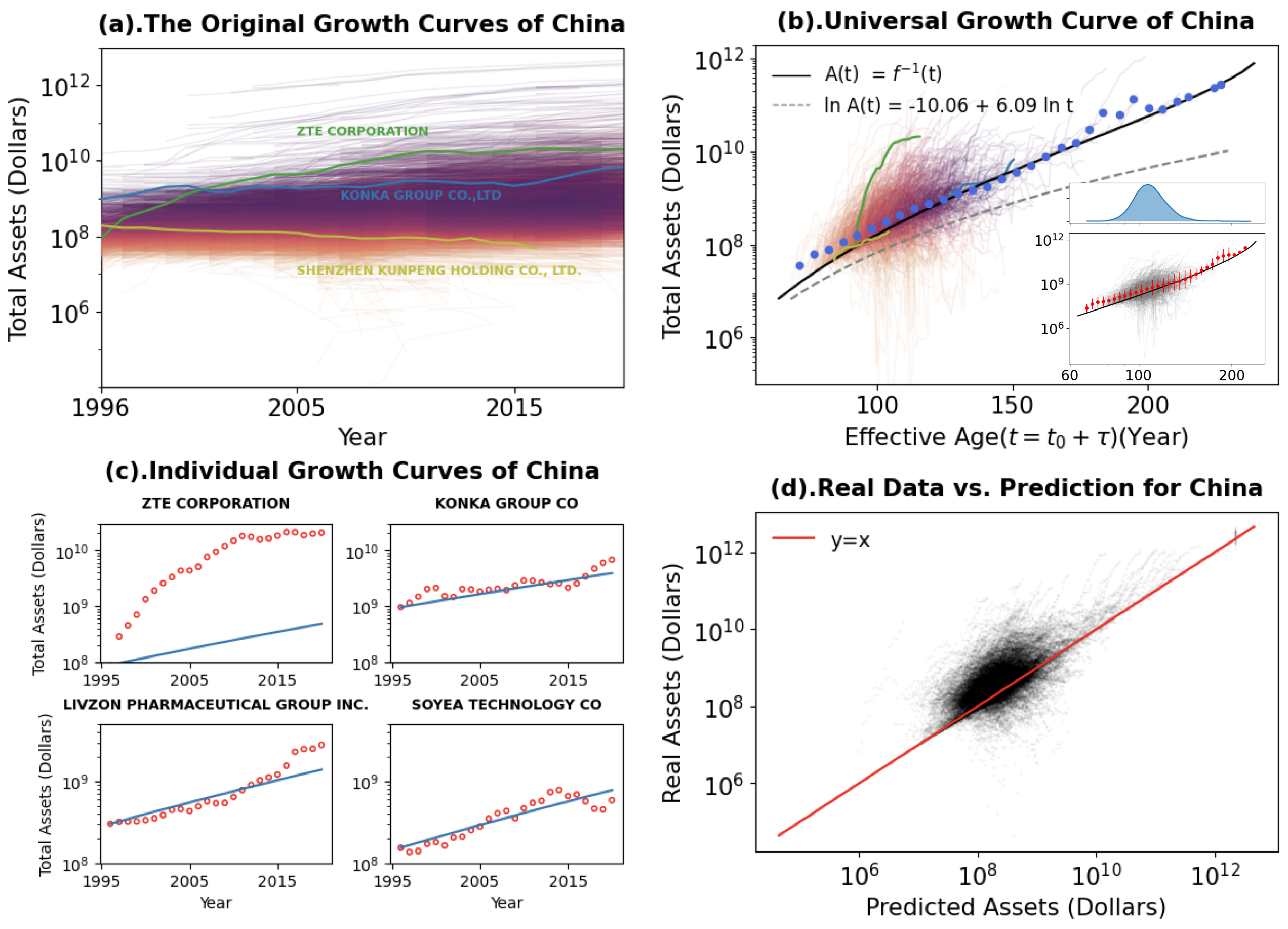}
\caption{\textbf{Publicly traded market of China}. The growth trajectories of all 3,160 companies in terms of total assets over time along with the generalized growth predictions.
(\textbf{a}) the raw growth trajectories of all companies in
our dataset from 1996 to 2020 along with selected companies representing over-performing, meeting, or under-performing predictions.
(\textbf{b}) The predicted generalized growth curves and the
translated growth trajectories of each company based in their $t_0$. We predict assets by Eq.~(\ref{eq:integratingequation}) from effective age, and approximately find the initial asset for each company. The inset shows the same graph on a log-log plot along with the distribution of companies.
(\textbf{c}) Observed growth trajectories (red circles) and their corresponding predicted growth (blue solid lines) obtained by solving Eq.~(\ref{eq:powerlawgrowth}) with the observed initial size for three selected companies. 
(\textbf{d}) The log-log plot of the predicted assets versus the observed assets in order to show the degree of the agreement between the theory and the data.
}
\label{fig:growthcurves_china}

\end{figure*}

The first important observation is that, like the US market, Chinese companies exhibit similar power law scaling in all of their major characteristics (see Fig.\ref{fig:scalings}). For the Chinese market, the exponent for net incomes is $\beta_I\approx 0.83$ which is very close to that of the US, $0.85$; however, the exponent for liability  is $\beta_L=1.09>1$, which is significantly different from that for the US. In this case, an explicit analytic expression for the generalized growth equation cannot be derived and we have to resort to a numerical solution for  inverting Eq.~\ref{eq:integratingequation} to obtain the growth curve. We find equally good predictions of the growth trajectory of Chinese companies as we did for US companies. Chinese companies display a similar diversity of company-level growth trajectories (Fig. \ref{fig:growthcurves_china}(a)) with the majority closely following the generalized growth curve and others significantly deviating (Fig.~\ref{fig:growthcurves_china}(b-c)), but the average prediction strongly agrees with data (Fig.~\ref{fig:growthcurves_china}(b,d)).

The growth equation can again be approximated by a power function of the same form as Eq.~\ref{eq:simplesolution2}. Here we find that $\gamma\approx 6.1$ (Fig. \ref{fig:growthcurves_china}(b), which is similar to the US where $\gamma=5.8$. This similarity arises because $\beta_I$, representing the scaling of net income, is nearly identical in the two markets. However, it is critical to recognize that the power-law approximation does not capture the data equally well in the US and China (see the dashed lines of Fig.~\ref{fig:growthcurves}(b) and Fig.~\ref{fig:growthcurves_china}(b)). This mismatch is because the character of the scaling of liability, captured by $\beta_L$, is radically different in these two systems. Specifically, in China $\beta_L$ is larger than one, and thus the approximation of Eq.~\ref{eq:simplesolution2} does not hold. This example illustrates that small differences in these exponents matter for the growth equation, affirming the predictive power of Eq.~\ref{eq:integratingequation} where the full form of the equation  captures the different characteristic of growth dynamics in China (Fig.~\ref{fig:growthcurves_china}). Even though the US and Chinese liability scaling exponents differ by only about $0.10$ - scaling linearly in the US but superlinearly in China -  this small difference leads to fundamental, and quite surprising differences in their growth, as revealed by our framework.

\subsection*{Distribution of Deviations}
Our generalized growth curve provides a natural baseline against which to quantify growth fluctuations. 
Although most companies conform well to the generalized growth behaviour, some companies deviate substantially. For example, Apple has grown faster than the prototypical growth curve, whereas American Plastics \& Chemicals has grown significantly slower. In contrast, Coca-Cola has followed the curve relatively closely, as illustrated in Fig.~\ref{fig:growthcurves}(b). 

To characterize how an individual company $i$ deviates from the generalized growth curve, we introduce
its average relative deviation over its lifespan, $T^i$:
\begin{equation}
\label{eq:error}
\epsilon^i=\sigma\left({\sum_{t=1}^{T^i}\epsilon^i(t)}\right)\frac{\sum_{t=1}^{T^i}\mid\epsilon^i(t)\mid}{T^i}
\end{equation}
Here $\sigma (x)$ is the sign function (i.e., $\sigma (x) = +1$ if $x>0$, and $-1$ if $x<0$); $\epsilon^i(t)\equiv \ln A^i(t)-\ln \hat{A}^i(t)\approx [{A^i(t)-\hat{A}^i(t)}]/{\hat{A}^i(t)}$ is the relative deviation of the predicted total assets at time step $t$, $A^i$ is the observed data, and $\hat{A}^i$ is the prediction. Note that by including the $\sigma$ function this measure gives the sign of the average deviation, either above or below the generalized growth curve, potentially providing a metric for a company's over- or under- growth performance relative to the expectation for its size. 

\begin{figure*}[ht!]
\center
\includegraphics[width=1.0\textwidth]{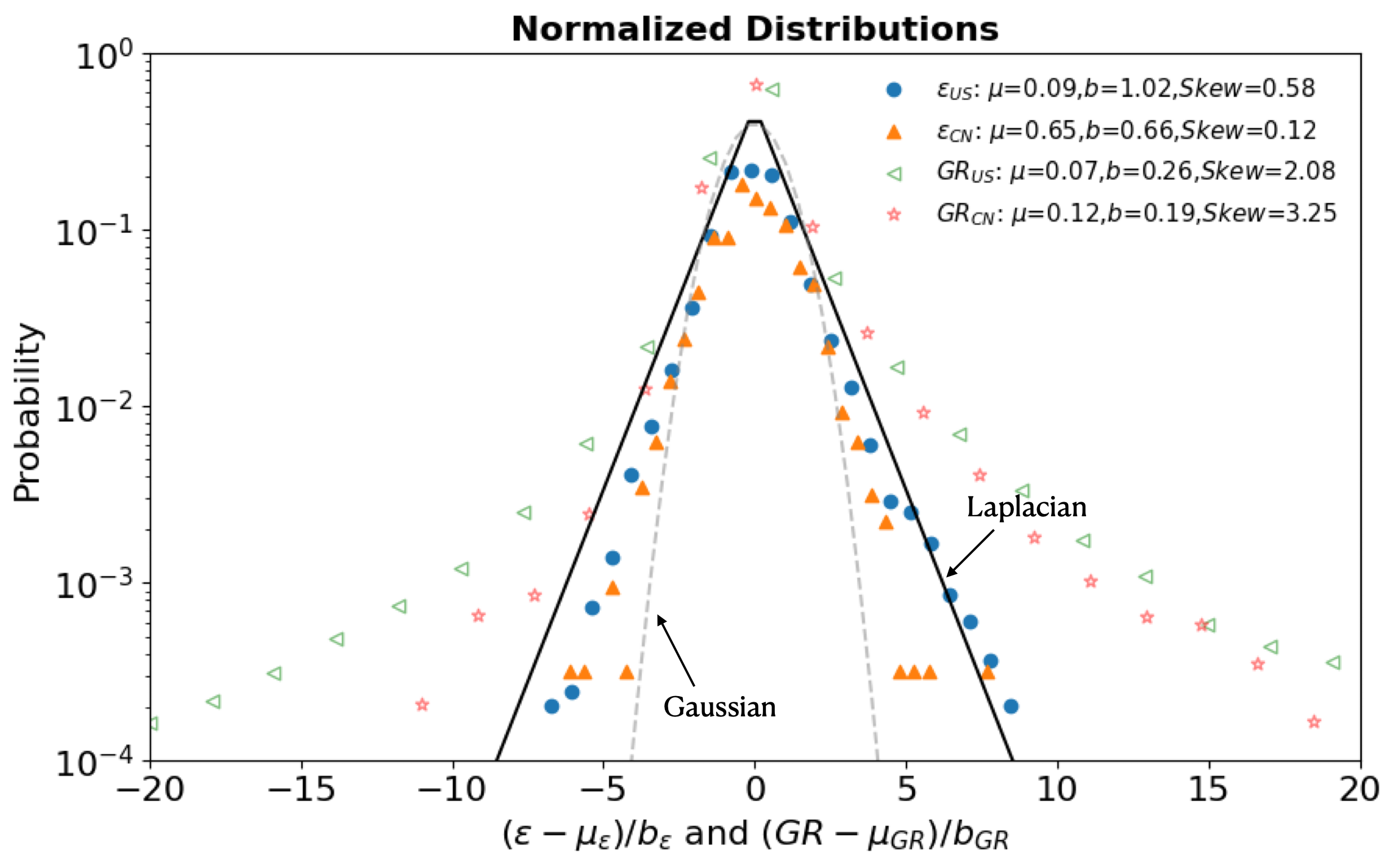}
\caption{The normalized distributions of deviations $\epsilon$ (filled circles or triangles) and growth rates ($GR\equiv \frac{A(t+1)-A(t)}{A(t)}\approx \ln A(t+1) - \ln A(t)$, empty triangles and stars) for the US (blue circles) and China (orange triangles). The black solid line is the standard Laplacian distribution ($\mu=0$ and $b=1$) and the gray dashed line is the standard Gaussian distribution. The parameters $\mu$ and $b$, as well as the skewness of the original Laplacian distributions are also given in the legend.}
\label{fig:errors}
\end{figure*}

The distributions of $\epsilon^i$ clearly exhibits a tent-like shape characteristic of a Laplace distribution as shown in Fig.~\ref{fig:errors}, defined by

\begin{equation}
    \label{eq:laplace}
    p(\epsilon)=\frac{1}{2b}\exp\left(\frac{-\mid\epsilon-\mu\mid}{b}\right),
\end{equation}
on the entire dataset for both US and China, where $\mu$ is the displacement parameter and $b$ the dispersion. We normalize $\epsilon$ by minus $\mu$ and divided by $b$ such that the distributions for both countries can be compared. We also plot the normalized distributions of the growth rates defined as $GR\equiv \frac{A^i(t+1)-A^i(t)}{A^i(t)}\approx \ln A^i(t+1)-\ln A^i(t)$ for all $t$ and $i$ because they are also reported to be laplacian distributed(e.g. \cite{stanley1996scaling,amaral1998power}).

Fig.~\ref{fig:errors} show the empirical distributions of all these quantities for both countries as well as the standard Laplace distribution and Normal distribution. We can see that the normalized distributions of $\epsilon^i$ for both US and China agree with the standard Laplacian distribution, while the growth rates in both countries can not. That means the growth rates are not exactly Laplacian distributed due to their large skewness on right. The underlying reason is the growth rates contain the averaged growth which is accounted by our generalized growth equations which are not pure fluctuations with symmetric distributions. These results help clarify previous proposals for Laplace distributions of fluctuations \cite{stanley1996scaling} and later discussions about about curvature away from such distributions \cite{perline2006volatility,fu2005growth}, by showing that it is necessary to normalize by our generalized growth curve before considering fluctuations in order to obtain a Laplace distribution. Such a procedure leads to a narrower Laplace distribution without serious curvature, but with significant skew.

It is also important to note that although different exponents of liability $\beta_L$ for US and China lead to different generalized growth curves with distinct power exponents, the deviations between the scaling laws and the generalized growth are the same, again illustrating the validity and generality of our method (Eq.~\ref{eq:powerlawgrowth} and Eq.~\ref{eq:generalgrowth}). In section S3.5 of SI, we also derive a relationship between the deviations of growth rates and scaling laws for both net incomes and liabilities which further confirms the tight connection between growth and scaling laws.
  
\subsection*{Growth, equity and debt}

The equity of a company is defined as the difference between its assets and liabilities: $Q \equiv A-L$. 
Thus, $dQ/dt = dA/dt - dL/dt$ which, from Eq.~(\ref{eq:balance 2}), is $I(t)$ so $dQ/dt = I(t)$, i.e., the rate of increase of equity is just the net income. Similarly, since $dQ/dA = 1 - dL/dA$, the growth equation (\ref{eq:generalgrowth}) can be expressed in a very compact form:
\begin{equation}
    \label{eq:equity}
    \frac{dA}{dt}=\frac{I(A)}{dQ/dA},
\end{equation}

Conceptually implicit in using power function scaling for both  $I(A)$ and $L(A)$ in Eq.~(\ref{eq:powerlawgrowth}) is that these are the prime quantities companies, on average, strive to optimise. In this sense, equity, $Q = A-L$, is a ``secondary'' derived quantity since the difference between two power functions cannot itself be a power function. Consequently, if equity is not directly optimised it cannot scale as a power function. It would therefore be technically incorrect to use power functions for both $I(A)$ and $Q(A)$ in Eq.~(\ref{eq:equity}), even though it might be a useful approximation. On the other hand, it is straightforward to consider the alternative scenario where equity is presumed to be the prime quantity companies try to optimise, rather than income and/or liabilities, and so obeys power function scaling. In that case, Eq.~(\ref{eq:simplesolution2}) is the {\it exact} solution for $A(t)$ with $\gamma = (\beta_Q - \beta_I)^{-1}$ and $c = {[(c_I/c_Q)(1 - \beta_I/\beta_Q)]}^{(\beta_Q - \beta_I)^{-1}}$, where $\beta_Q$ and $c_Q$ are the scaling exponent and normalization constant of $Q$ with respect to $A$, respectively.

It is instructive to consider the realistic case when $\beta_L\approx 1$, which holds for the US market as a whole. In that case, $c_L = k$, the  approximately constant debt ratio: $k = L(A)/A$ and $Q(A) = (1 - k)A$ (i.e., $\beta_Q = 1$ and $c_Q = 1 -k$). Consequently, the solution for $A(t)$ reduces precisely to Eq.~(\ref{eq:simplesolution}).
The ratio $A/Q$ is the financial leverage ratio conventionally referred to as the equity multiplier, 
$r$ [$ = (1 - k)^{-1}$], in terms of which Eq. (\ref{eq:equity}) becomes ${dA}/{dt}\approx rI$. That is, the growth rate of a company is its net income modulated by the equity multiplier leverage ratio.

Note that when the equity multiplier,  $r=1, k=0$, so $L = 0$ and $Q = A$ and all assets owned by a company are held in stockholder equity and none are funded by debt, in which case company growth is determined solely by capital in the form of net income. In our data we find $r\approx 2$, so $k\approx 0.5$ implying that, on average, about 50\% of a company's equity is funded by debt, or equivalently, that the total assets of the average company are twice those held in stockholder equity. Perhaps counter-intuitively, this implies that, holding all else constant, issuing debt increases growth by magnifying the returns from net income (i.e., retained earnings): debt is an additional inflow of finances thereby increasing the total assets of a company, and because growth is a function of assets, companies with debt will grow faster than those without. On the other hand, issuing debt incurs risks and costs; higher debt potentially leads to increased debt expenses, higher leverage, lower flexibility, lower investor confidence, and the greater risk of bankruptcy. However, at low interest rates debt compensates for short term losses and provides additional capital to finance growth where the additional interest costs can be offset by higher returns on capital invested in the company. Consequently, the ability to manage debt and its associated risk is a crucial factor in the long-term growth of a company.

\begin{figure*}[ht!]
\includegraphics[width=1.0\textwidth]{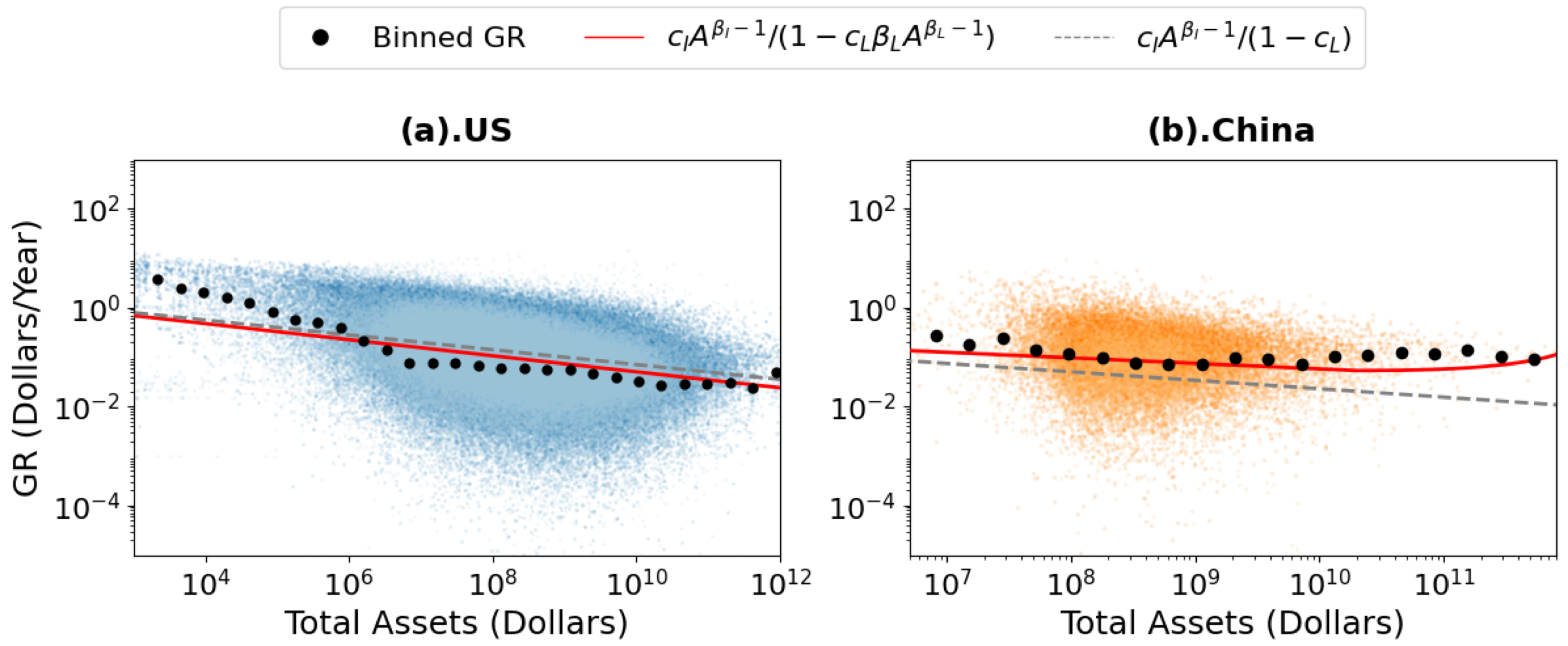}
\caption{The dependence of the growth rates on total assets for both the US ({\textbf{a}}) and China ({\textbf{b}}) over all available years along with the theoretical predictions. For the two figures, both the positive (darker colored) and the negative growth rates (lighter colored) are shown together, on which, the absolute values of the negative growth rates are plotted to show them on the logarithmic scale. To clarify the relationship between the growth rates and the total assets for the data, we divided the total assets of companies into logarithmic (30 for America and 20 for China) bins and calculated the average growth rate in each bin. These averages are plotted as the black data points. The red lines are the theoretical predictions from our model $dA/(A\cdot dt)=c_I\cdot A^{\beta_I-1}/(1-c_L\cdot \beta_L\cdot A^{\beta_L-1})$, and the dashed grey line are the approximations according to $dA/(A\cdot dt)\approx c_I\cdot A^{\beta_I-1}/(1-c_L)=r\cdot c_I\cdot A^{\beta_I-1}$ when $\beta_L\approx 1$.}
\label{fig:gibrat}
\end{figure*}

It follows that the asset growth rate is the return on equity - the ratio of net income to stockholder equity (net assets) - which is a measure of the ability of a company to generate profits from the assets it holds;
\begin{equation}
\label{eq:growthrate}
GR=\frac{c_IA^{\beta_I-1}}{1-c_L\beta_LA^{\beta_L-1}}\stackrel{\beta_L\rightarrow 1}\longrightarrow r \frac{I}{A}=\frac{I}{Q}=r c_I A^{\beta_I-1}.    
\end{equation}
Because $\beta_I<1$, Eq.(\ref{eq:growthrate}) captures the decreasing returns on net assets with increasing company size. This prediction is quantitatively confirmed by data as shown in Fig. \ref{fig:gibrat} for both countries and shows explicitly why Gibrat's assumption of a constant relative growth rate is incorrect.

\section{Discussion}\label{sec12}
Here we derived a general equation for the growth of companies, which connects scaling exponents to  diverse possibilities for the form of growth. Overall our model does an excellent job of predicting the growth of US companies, and reveals new generalities in both growth trajectories and their fluctuations. Impressively, this framework works equally well for Chinese companies, where it is important to note that both the scaling exponents and subsequent form of the growth trajectory significantly differ from the US. Application to this second and very different market demonstrates the generality and flexibility of our theoretical framework.

More broadly, our theory is similar in spirit to the dynamics of growth processes of other entities well-described by scaling phenomena, such as populations, cities, nations, mammals, plants, or bacterial colonies \cite{west2017scale}. Our growth equation is part of a general class of models which captures how the scaling of inputs and outputs determines the specific form of growth across this diverse range of entities and institutions. For example, in mammals, \emph{income} is the total available metabolic energy while \emph{costs} are the energy required for repair and maintenance, both of which scale systematically with body size. Because the latter scales more quickly than metabolic rate  - the exponents are 1 and 3/4 respectively - body size reaches a maximum  where maintenance consumes all of the incoming energy and growth ceases, leading to a stable adult size \cite{west2001general}. There is no indication of a similar maximum size in companies based on the growth curves or on the scaling laws for liabilities and net income which predict the observed open-ended power-law growth in time. Another example is cities whose metabolism scales superlinearly with population (i.e., with an exponent greater than 1) whereas its per-capita costs are approximately constant \cite{bettencourt2007growth}. In this scenario growth is open-ended as it is for companies; however, rather than growing as a power function in time, cities grow faster than exponentially (\emph{super}exponentially) leading to a finite time singularity. Bacteria exhibit a similar scaling pattern to cities, namely,  superlinear ``incomes'' and linear costs \cite{delong2010shifts,kempes2012growth}. Companies are yet another variation on this theme in which the sublinear scaling of net income and the market-dependent linear or sublinear scaling of liabilities with assets leads to \emph{sub}exponential growth. Consequentially, companies are open-ended growers, although other factors may limit their survival and lifespan \cite{cefis2021understanding}. Their size is dependent on their ability to deploy assets in response to external market conditions, rather than pre-determined by a genetic code, such as in the ontogeny of mammals. However, the ability of a company to deploy assets to promote growth is size-dependent.

A mechanism seemingly unique to companies and not found in other complex adaptive systems is the ability to raise capital from markets by issuing liability or equity to manage growth. While there may be biological precursors to offsetting immediate growth demands by either borrowing resources from kin for mutual benefit (i.e., gestation, followed by survival and eventual reproduction), or buffering against future uncertainty (i.e., storage, hibernation, or estivation, for example), arguably, the ability to raise capital by selling somatic control (i.e., stockholder equity), or by borrowing against perceived future value (i.e., debt) is unique to, and perhaps definitive of, markets. Debt is a fundamental component in the evolution of human economies, as it is a mechanism that allows for economic transactions to be deferred and to remain incomplete for a finite period of time through negotiation \cite{graeber2012debt}. This ability to raise capital by borrowing from the future means the growth trajectories of companies are, in many ways, more flexible than organisms or cities, as runs of bad years can be temporarily mitigated. However, capital investments are not free and so this mitigation can only offset, not replace, productivity \cite{roberts2007modern}. Moreover, this has interesting implications for theories of the firm: not only do companies reduce transaction costs, aggregate information, and achieve economies of scale, they also provide mechanisms for buffering by being able to displace immediate shortfalls into the future. This buffering is much harder to sustain on an individual level, especially as the primary mechanism for storing and building wealth (and other buffering mechanisms) is the market itself. Public companies and markets only exist through flows of finance with the expectation of future growth, and the ability of public companies to perform this role ultimately determines their fate.

The unique power-law growth with time we find here has its origins in the specific scaling of financial quantities in companies, where liabilities, on average, scale with assets with an exponent consistent with unity. This approximate linearity implies that debt is independent of size, and so the liabilities accessible per unit of assets does not depend on the total assets of the company \cite{modigliani1958cost}. Note, however, that for small companies there are deviations from this, as reflected in the curvature away from the power function at the lower end of the plots in Fig.~(\ref{fig:scalings}) and the inset of Fig.~\ref{fig:growthcurves}(b).

These results highlight the importance of the values of the exponents for net income, $\beta_I$, and liability $\beta_L$, since both of them can affect growth. The exponent of liability, $\beta_L$, can affect the shape of the growth curve as shown for Chinese companies. However, when $\beta_L\approx 1$, from Eqs.~(\ref{eq:simplesolution}) and (\ref{eq:simplesolution2}), the time exponent governing growth, $\gamma \approx 1/(1-\beta_I)$. Consequently, small changes in $\beta_I$ have potentially large effects on $\gamma$ and therefore on growth trajectories. For instance, if $\beta_I$ were 1, and not 0.85, then companies would grow exponentially in time, whereas an exponent of $0.9$ would give a temporal scaling of $t^{10}$, while 0.8 would give $t^5$. The value of the overall normalisation constant, $c$, is equally sensitive to small changes in the scaling parameters.  The sensitivity of both $\gamma$ and $c$ is closely related to the well-known sensitivity of marginal returns in determining firm survival in general. 
Developing mechanistic theory to explain why we see these exponents will be of considerable importance given the consequences for understanding scaling of company growth over time. In both biological systems and cities scaling exponents have their origin in  the dynamics and geometry of internal network structures that transport energy, resources and information, such as vasculature, roads, and social networks.  For firms there have been several proposals of how internal structure leads to growth dynamics \cite{amaral1998power,stanley1996scaling, fu2005growth}. While these models do not predict scaling exponents nor the growth dynamics we predict and observe, perhaps similar inherent constraints of company structure lead to the sublinear scaling we report. Additionally, the extreme competitiveness of markets at short time scales and how larger firms \emph{feel the size of the market} also constrain internal function and these effects of the overall \emph{ecology} could likely be the driver of the sublinear scaling. 

Importantly, Gibratian growth is the special case of the theory we develop here, in which net income scales linearly with assets leading to purely exponential growth. However, the inherently nonlinear scaling of net income reveals that company growth cannot be a Gibrat process. This result also suggests a paradigmatic shift in how we should think about companies grow over time: In exponential systems, we are conditioned to think about compound interest and growth as a constant proportion of size, and thus exponential in time. However, we have shown that the expected growth rate of a company is a sublinear function of its size, and so {\it the capacity for growth decreases as a company increases in size}. As a company becomes larger a decreasing proportion of its assets can be deployed to promote growth: The more assets a company deploys toward maintaining market share and meeting production, the less flexible it is in responding to external fluctuations or diversifying it's production \cite{roberts2007modern}.

It is interesting to note that while the ensemble of companies grows following a power function of time, the overall market grows exponentially. In fact, an average company will achieve their fastest rate of growth at intermediate times out-performing the average growth of the market, but only for a finite period. Over this period, companies may either merge or be acquired by competitors, or eventually succumb to mounting maintenance costs. An open question then is why the market grows exponentially while the ensemble average of companies does not. The answer is likely to do with the constant and rapid turnover of companies entering and exiting markets through birth and death processes. Here it should be noted that while other regularities are known for the death dynamics of companies \cite{daepp2015mortality} the framework presented here does not predict market entry and exit dynamics, only the form of growth. It will be important for future work to connect growth with exit probabilities and to the growth dynamics of firms prior to the initial public offering or to private companies.

\section{Methods}\label{secA1}

\subsection*{Dataset}
\label{sec:dataset}

\subsubsection*{(a). US}
\label{sec:dataset_US}
American dataset contains the financial information obtained from the income statements and balance sheets of publicly traded companies from 1950 to 2018. There are 31,553 companies in total and most of them are from North American and overseas American Depository Receipt firms. The data is from Compustat North America and Compustat Historical databases compiled by Standard \& Poor's\cite{SP}. 

The classification for companies is according to the North American Industry Classification System(NAICS) standard which is developed by the statistical agencies of Canada, Mexico and the United States. A six digits code is assigned to each company except 1,839 companies. We classify the companies into 447 economic sectors according to the code. The list of the sectors are provided in S7 of SI.

\subsubsection*{(b). China}
\label{sec:dataset_CN}
Chinese dataset contains the financial information obtained from the income statements and balance sheets of publicly traded companies from 1996 to 2020. There are 3,160 companies in total.

\subsection*{Deflation}
\label{sec:deflation}
Most financial variables are measured by U.S. dollars, therefore the inflation factor should be considered. Especially when we compare the same indicator for different years, the inflation factor plays an important role. In this paper, we deflate each financial variable in the same way. Take sales as an example. At first, we select the last year as the base year $T$, then all the sales data will be converted and measured by the money in year $T$ as shown in equation \ref{eq:s1}.
\begin{equation}
\label{eq:s1}
    S(t)=S_0(t)e^{r(t)}
\end{equation}
$S(t)$ is the deflated sales at year $t$, $S_0(t)$ is the raw data of sales at $t$. $r(t)$ is the inflation rate relative to $T$, it can be computed as:
\begin{equation}
    r(t)=\sum_{\tau=1}^{T}\pi_{\tau}
\end{equation}
Where, $\pi_{\tau}$ is the inflation rate of time $\tau$ to the previous year. Thus, all the values shown in the main text are normalized in this way.

\subsection*{Fitting Methods}
\label{sec:fitting}
In the main text, we use two different methods to fit the data on log-log coordinate to obtain the scaling laws. For the 423 individual economic sectors, we used the linear mixed-effect models (LMMs) to fit and estimated the exponents and coefficients of the scalings. For the data set of all sectors mixed together, we at first flat all the data points for all companies in all years together, and we used the ordinary least square (OLS) method to fit the data and estimated parameters. The comparing results of the two methods on estimating scaling exponents and coefficients are shown in Section S2.1 of the SI.

LMMs are extensions of linear regression models for data that are collected in groups. The assumption behind LMM is that the coefficients can vary with respect to one or more grouping variables. In our case, for any company within the sector that we considered, and for any aggregated-level variable $X$, we assumed that each data point $(\ln A^{t,i}, \ln X^{t,i})$ for company $i$ at year $t$ follows a scaling law, but the scaling exponent and coefficient vary with the company $i$. That is,
\begin{equation}
    \ln X^{t,i}=\ln(c_X^{i})+\beta_X^{i}\ln A^{t,i}+\xi^{t,i} 
\end{equation}
for the company $i$ at time $t$, where, $\xi^{t,i}\sim N(0,\sigma^2)$ is an independent random number, and $\sigma$ is a constant.  And $\ln(c_X^{i}), \beta_X^{i}$ are the company dependent intercept and slope respectively, and satisfy:
\begin{equation}
    \ln(c_X^i)=\ln(c_X)+\xi_c^i,
\end{equation}
and:
\begin{equation}
    \beta_X^i=\beta_X+\xi_{\beta}^i,
\end{equation}
where, $\xi_c^i\sim N(0,\sigma_0^2)$ and $\xi_{\beta}^i\sim N(0,\sigma_1^2)$, and $\sigma_0,\sigma_1$ are constants. We then fit the linear mixed-effect model to obtain the fixed effect exponent $\beta_X$ and the coefficient $c_X$ as the estimations. 

However, we give up using the LMM method to estimate the exponent and the coefficient for all sectors. The reason is that because the exponent will not be used for estimating the relation between $X$ and $A$ but also for deriving the growth equation, and to better fit the data through time, we adopt the OLS method to estimate the exponent and the coefficient. 

\subsection*{Definition of Natural Age}
\label{sec:natural_ages}
The first year a company $i$ is born is not reported in our data, and so we treat the first year that the assets of company $i$ appears in the data as it's first year. We define the natural age of $i$ as the time span from its first year to the current year (i.e., the number of years $i$ appears in the data). 
\subsection*{Estimation for the leverage $r$}
Because $\beta_L\approx 1$ for the whole market, therefore $L\approx k\cdot A$, where $k\approx c_L$. Since $Q=A-L$, thus, $Q\approx r\cdot A=(1-k)\cdot A$.

To estimate $k$ for all companies, we cannot use $c_L$ estimated by equation \ref{eq:scaling} directly because this is not a linear relationship. We use equation \ref{eq:r_cal} instead of algebraic average value to estimate because the size range is very large.
\begin{equation}
\label{eq:r_cal}
    k=\exp\left[\langle \ln{L}-\ln{A}\rangle\right],
\end{equation}
where the average $\langle\cdot\rangle$ is taken over all data points. This is equivalent to the OLS estimation of $k$ by taking the slope is exactly 1 from the relationship $\ln L=\ln k+\ln A$. Therefore, $r$ is calculated by $1-k$

\section{Figures}\label{sec6}

\begin{table*}[ht!]
\center
\small
\caption{The scaling exponents $\beta_X$ and the constants $c_X$(logarithm) with the $95\%$ CIs for selected variables in two countries. Other statistics of the OLS fits are reported in Section S2.1 of the SI.}
\label{tab:scaling}
\begin{tabular}{| c|c|c|c|c|}
\hline
Variables&\multicolumn{2}{|c|}{US} &\multicolumn{2}{|c|}{China} \\
\hline
-&Exponent & $\ln$(Constant) &Exponent&$\ln$(Constant)\\
\hline
\multicolumn{5}{|c|}{Income related} \\
\hline
Sales&0.90[0.90,0.90]&1.53[1.5,1.57]&0.92[0.91,0.92]&0.93[0.84,0.84]\\
\hline
Net Incomes&0.85[0.85,0.85]&-0.19[-0.23,-0.15]&0.83[0.82,0.84]&0.09[-0.04,-0.04]\\
\hline
EBITDA&0.94[0.93,0.94]&-1.03[-1.06,-1.0]&0.87[0.86,0.87]&-0.30[-0.41,-0.41]\\
\hline
Gross Profit&0.85[0.85,0.85]&1.45[1.42,1.48]&0.91[0.9,0.91]&3.0[2.74,3.29]\\
\hline
Dividends&0.56[0.55,0.58]&5.3[4.94,5.65]&0.79[0.78,0.80]&-0.06[-0.23,-0.23]\\
\hline
Retained Earnings&0.90[0.90,0.90]&0.17[0.13,0.22]&0.96[0.95,0.96]&-1.51[-1.65,-1.65]\\
\hline
\multicolumn{5}{|c|}{Cost related}\\
\hline
Cost of Sales&0.85[0.85,0.85]&2.01[1.98,2.05]&0.91[0.90,0.91]&1.10[1.01,1.01]\\
\hline
Total Tax&0.93[0.93,0.94]&-2.71[-2.76,-2.65]&1.03[1.03,1.04]&-6.34[-6.48,-6.48]\\
\hline
Operating costs&0.79[0.79,0.79]&3.55[3.52,3.58]&1.02[1.01,1.03]&-1.67[-1.9,-1.9]\\
\hline
R\&D&0.70[0.7,0.70]&2.48[2.42,2.54]&0.61[0.58,0.63]&3.48[2.98,2.98]\\
\hline
\multicolumn{5}{|c|}{Size related}\\
\hline
Employee&0.74[0.73,0.74]&-7.7[-7.74,-7.67]&0.63[0.62,0.64&-6.95[-7.07,-7.07]\\
\hline
Cash&0.82[0.82,0.82]&0.14[0.1,0.18]&0.99[0.99,1.0]&-1.92[-2.04,-2.04]\\
\hline
Total liabilities&0.99[0.99,0.99]&-0.46[-0.48,-0.44]&1.09[1.09,1.10]&-2.77[-2.84,-2.84]\\
\hline
\end{tabular}
\end{table*}

\end{document}